\documentclass[entropy,accept,moreauthors,pdftex,12pt,a4paper]{mdpired} 
\usepackage{bbm,graphicx,subfigure,color}

\newcommand{\rr}{\mathbbm{R}}
\newcommand{\red}[1]{#1}

\newcommand{\id}{\mathbbm{1}}
\renewcommand{\text}[1]{{\rm #1}}
\newcommand{\hh}{\mathcal{H}}
\newcommand{\T}{^{\sf T}}
\newcommand{\gr}[1]{\underline{#1}}

\newcommand{\eq}[1]{Equation~(\ref{#1})}

\newcommand{\ket}[1]{|#1\rangle}
\newcommand{\bra}[1]{\langle#1|}

\newcommand{\op}[1]{\hat{#1}}
\newcommand{\adj}[1]{{#1}^{\dag}}
\newcommand{\comm}[2]{\left[#1,#2\right]}
\newcommand{\sig}{{\boldsymbol\sigma}}
\newcommand{\gam}{\boldsymbol{\gamma}}

\lastpage{12}
\pubvolume{}
\pubyear{2013}
\history{ \quad \\ \quad }
\pdfoutput=1
\Title{Genuine Tripartite Entanglement and Nonlocality in Bose-Einstein Condensates by Collective Atomic Recoil}

\Author{Samanta Piano $^{1}$ and Gerardo Adesso $^{2,}$*}

\address{%
$^{1}$ Midlands Ultracold Atom Research Centre, School of Physics and Astronomy, The University of Nottingham, Nottingham NG7 2RD, UK; E-Mail: samanta.piano@nottingham.ac.uk\\
$^{2}$ School of Mathematical Sciences, The University of Nottingham, Nottingham NG7 2RD, UK}

\corres{E-Mail:~gerardo.adesso@nottingham.ac.uk; Tel.: +44-0-115-84-66165.}

\abstract{ We study a system represented by a Bose-Einstein condensate interacting with a cavity field in presence of a strong off-resonant pumping laser. This system can be described by a three-mode Gaussian state, where two are the atomic modes corresponding to atoms populating upper and lower momentum sidebands and the third mode describes the scattered cavity field light. We show that, as a consequence of the collective atomic recoil instability, these modes possess a genuine tripartite entanglement that increases unboundedly with the evolution time and is larger than the bipartite entanglement in any reduced two-mode bipartition. We further show that the state of the system exhibits genuine tripartite nonlocality, which can be revealed by a robust violation of the Svetlichny inequality when performing displaced parity measurements. Our exact results are obtained by exploiting the powerful machinery of phase-space informational measures for Gaussian states, which we briefly review in the opening sections of the paper.
}


\begin{document}

\section{Introduction}

Quantum information science is a fast-rising area of investigation that capitalizes interest from a broad spectrum of disciplines, including physics, mathematics, computer science, biology, and philosophy \cite{nichu}. In particular, the field of atomic and optical physics is experiencing a fruitful cross-fertilization with quantum information research. On one hand, tools developed for the study of quantum informational concepts such as entanglement \cite{entanglement} have proven useful to simulate efficiently many-body dynamics~\cite{peps}. On the other hand, atomic, optical systems, and their interfaces, offer valuable test grounds to implement prototypical protocols in quantum information, communication, computation and metrology~\cite{qinternet}. The branch of continuous variable quantum information, which exploits quantum correlations between continuous degrees of freedom such as light quadratures or collective spin components of atomic ensembles, is an especially versatile one for the characterization of quantum tasks \cite{brareview}. In particular, Gaussian states of continuous variable systems constitute a precious resource for theoretical studies of quantum correlations \cite{ourreview} and for unconditional implementation of quantum information processing \cite{furuscience}. Gaussian states naturally occur as ground or thermal equilibrium states of any physical quantum system in the ``small-oscillations'' limit, and can be very efficiently engineered, controlled and detected in various laboratory setups, including light, ultracold atomic ensembles, trapped ions, nano-/opto-mechanical resonators, and hybrid interfaces thereof \cite{book}.

Despite such an important role played by Gaussian states, a comprehensive and satisfactory characterization of their informational properties, including entanglement and more general types of quantum correlations, has been accomplished only very recently \cite{ourreview,pirandolareview}. In this work, we shall recall the basic notions of Gaussian quantum information and correlations using a self-contained formalism based on the R\'enyi entropy of order $2$ \cite{renyi}. We will then apply the presented toolbox to study and quantify genuine tripartite entanglement in a particular system, composed by a Bose-Einstein condensate coupled with a single-mode quantized field in a ring cavity and driven by a strong far off-resonant pump laser~\cite{bonifacio,pariscola,pariscoladissipative,EXPCARL}.

Astonishing progress in the manipulation and control of quantum systems at the mesoscopic scale has enabled in recent years \red{the design and implementation of groundbreaking experiments} involving cold atoms in cavity-generated dynamical optical potentials \cite{2013}. In the system we analyze, in the linear regime around the equilibrium momentum state of the atoms, an effective interaction between the cavity mode and two atomic side momentum modes is realized. Under these approximations, the state of the dynamical system is a three-mode Gaussian state \cite{pariscola,pariscoladissipative}. Applying our formalism, in the limit of a lossless cavity, a strong tripartite entanglement generated by collective atomic recoil is found in all regimes of operation. We also discuss the quantumness of correlations \cite{zurek} in the various atom-atom and atom-field partitions. Finally, we show that the studied system allows for a violation of the Svetlichny inequality \cite{svelticchio}, revealing genuine tripartite nonlocality in a broad parameter range. The latter can be detected experimentally by displaced parity measurements on the three-mode system \cite{wodk,Mauro_Referee,maurosvelt}.

The paper is organized as follows. In Section~\ref{secG} we review definitions and informational measures for Gaussian states. In Section~\ref{secB} we describe the driven intracavity ultracold atomic system. In Section~\ref{secR} we present our results on quantum correlations and nonlocality of the considered system. Section~\ref{secC} summarizes and concludes the manuscript.

\newpage
\section{The Preliminary Toolbox: Quantum Correlations of Gaussian States}\label{secG}

A continuous variable system of $N$ canonical bosonic modes is described by a Hilbert space \linebreak $\hh=\bigotimes_{k=1}^{N} \hh_{k}$
resulting from the tensor product structure of infinite-dimensional Fock spaces $\hh_{k}$'s, each of them associated to a single mode \cite{eisertplenio,brareview,ourreview}. For instance, one can think of \red{the non-interacting quantized electromagnetic field}, whose Hamiltonian $\op{H} = \sum_{k=1}^N \hbar \omega_k \left(\op{n}_k +
\frac12\right)$
describes a system of an arbitrary number $N$ of harmonic oscillators of different frequencies, the {\em modes} of the field. Here $\op{n}_k=\adj{\op{b}}_k\op{b}_k$ is the number operator for mode $k$, with $\op{b}_k$ and $\adj{\op{b}}_k$ being respectively the annihilation and creation operators of an excitation in mode $k$ (with frequency $\omega_k$), which satisfy the bosonic commutation~relation
\begin{equation}\label{CV:comm}
\comm{\op{b}_k}{\adj{\op{b}}_{k'}}=\delta_{kk'}\,,\quad
\comm{\op{b}_k}{\op{b}_{k'}}=\comm{\adj{\op{b}}_k}{\adj{\op{b}}_{k'}}=0\,
\end{equation}

In the following, when not explicitly stated otherwise, we shall adopt natural units with $\hbar=c=1$, where $\hbar$ is the reduced Planck constant and $c$ is the speed of light. The corresponding quadrature phase operators (``position'' and ``momentum'') for each mode are defined as $\hat q_{k} = \frac{(\op b_{k}+\op b^{\dag}_{k})}{\sqrt{2}}$ and $\hat p_{k} = \frac{(\op b_{k}-\op b^{\dag}_{k})}{i \sqrt 2}$. We can group together the canonical operators in the vector $\gr{\hat R}=(\hat{q}_1,\hat{p}_1,\ldots,\hat{q}_N,\hat{p}_N)\T
\in \mathbbm{R}^{2N}$, which enables us to write in compact form the bosonic commutation relations between the quadrature phase operators, $[\hat{R}_k,\hat{R}_l]= i\Omega_{kl}$, where $\boldsymbol\Omega$ is the $N$-mode symplectic form
\begin{equation}
\boldsymbol\Omega=\bigoplus_{k=1}^{N}\boldsymbol\omega\, , \quad \boldsymbol\omega=
\left(\begin{array}{cc}
0&1\\
-1&0
\end{array}\right)\, \label{symform}
\end{equation}

Gaussian states, such as coherent, squeezed and thermal states, are completely specified by the first and second statistical moments of the phase quadrature operators. As the first moments can be adjusted by marginal displacements, which do not affect any informational property of the considered states, we shall assume them to be zero, $\langle \gr{\hat{R}}\rangle=0$ in all the considered states without loss of generality. The important object encoding all the relevant properties of a Gaussian state ${\op{\varrho}}$ is therefore the covariance matrix (CM) $\sig$ of the second moments, whose elements are given by
\begin{equation}\label{sigmaij}
(\sig)_{j,k}=\text{tr}[{\op{\varrho}} \{\hat{R}_j,\hat{R}_k\}]\,
\end{equation}

Any undisplaced $N$-mode Gaussian state with CM $\sig$ can be equivalently described by a positive, Gaussian Wigner phase-space distribution of the form
\begin{equation}
\label{eq:wigner}
W_{\sig}(\gr{\xi}) = \frac{1}{\pi^N \sqrt{\det\left(\frac\sig2\right)}} \exp\left[-\gr{\xi}^{\sf T} \left(\frac\sig2\right)^{-1} \gr{\xi}\right]\,
\end{equation}
with $\gr{\xi} \in \rr^{2N}$ a phase-space vector.

An extensive account of informational and entanglement properties of Gaussian states, using various well-established measures, can be found for instance in \cite{eisertplenio,ourreview,pirandolareview}. Here we follow a very recent comprehensive approach introduced in~\cite{renyi}, to which the reader is referred for further details and rigorous proofs.

R\'{e}nyi-$\alpha$ entropies are a family of additive entropies, which provide a generalized spectrum of measures of (lack of) information in a quantum state ${\op{\varrho}}$. They are defined as
\begin{equation}\label{eq:ren}
{H}_\alpha({\op{\varrho}}) = \frac1{1-\alpha} \ln \text{tr} ({\op{\varrho}}^\alpha)\,
\end{equation}
and reduce to the conventional von Neumann entropy in the limit $\alpha \rightarrow 1$. The case $\alpha=2$ is especially simple, $H_2({\op{\varrho}}) = - \ln \text{tr} ({\op{\varrho}}^2)\,.$
For arbitrary Gaussian states, the R\'enyi entropy of order $2$ satisfies the strong subadditivity inequality \cite{renyi}; this makes it possible to define relevant Gaussian measures of information and correlation quantities, encompassing entanglement and more general quantum and classical correlations, under a unified approach. We will then adopt $H_2\equiv H$ (omitting the subscript ``$2$'' from now on) as our preferred measure of mixedness (lack of purity, or, equivalently, lack of information, \emph{i.e}., ignorance) for a generic Gaussian state ${\op{\varrho}}$ with CM $\sig$. Explicitly,
\begin{equation}\label{eq:renyg}
H(\sig) = \frac12 \ln (\det \boldsymbol{\sigma})\,
\end{equation}
which vanishes on pure states ($\det \boldsymbol{\sigma}^{\rm pure}=1$) and grows unboundedly with increasing mixedness of the state. This measure is directly related to the phase-space Shannon entropy of the Wigner distribution $W_{\sig}$ of the Gaussian state with CM $\sig$, Equation~(\ref{eq:wigner}), sampled by homodyne detections \cite{renyi}.

A measure of bipartite entanglement $E$ \cite{entanglement} for Gaussian states based on R\'enyi-$2$ entropy can be defined as follows \cite{renyi}. Given a Gaussian state ${\op{\varrho}}_{AB}$ with CM $\sig_{AB}$, we have
\begin{equation}\label{eq:GR2_ent}
E_{A|B} (\sig_{AB}) = \inf_{\left\{\gam_{AB}\ :\ 0<\gam_{AB} \le \sig_{AB}, \, \det{\gam_{AB}}=1\right\}} \frac12 \ln \left(\det \gam_A\right)\,
\end{equation}
where the minimization is over pure $N$-mode Gaussian states with CM $\gam_{AB}$ smaller than $\sig_{AB}$. For a pure Gaussian state ${\op{\varrho}}_{AB} = \ket{\psi_{AB}}\bra{\psi_{AB}}$ with CM $\sig_{AB}^{\rm pure}$, the minimum is saturated by $\gam_{AB}=\sig_{AB}^{\rm pure}$, so that the measure of \eq{eq:GR2_ent} reduces to the pure-state R\'enyi-$2$ entropy of entanglement,
\begin{equation}\label{eq:GR2_ent_pure}
E_{A|B}(\sig_{AB}^{\rm pure})= H(\sig_A) = \frac12 \ln (\det{\sig_A})\,
\end{equation}
where $\sig_A$ is the reduced CM of subsystem $A$. For a generally mixed state, Equation~(\ref{eq:GR2_ent}) amounts to taking the Gaussian convex roof of the pure-state R\'{e}nyi-$2$ entropy of entanglement, according to the formalism of \cite{geof,ourreview}. Closed formulae for $E$ can be obtained for special classes of two-mode Gaussian states \cite{renyi}, including marginal partitions from three-mode pure Gaussian states.

For pure states, entanglement is the only type of quantum correlation. For mixed states, even most separable states can display nonclassical features in their correlations \cite{modireview}. Quantum discord \cite{zurek,vedral} captures such features by quantifying the minimum informational disturbance induced on the state of a bipartite system by performing a local measurement on one of the subsystems. A Gaussian measure of discord \cite{giordaparis,adessodatta,renyi} based on R\'enyi-$2$ entropy can be defined for a bipartite Gaussian state ${\op{\varrho}}_{AB}$ as the difference between total and classical correlations (see e.g.,~\cite{renyi} for details), minimized over local Gaussian measurement. Depending on the subsystem on which the measurement is performed (\emph{i.e}., the subsystem whose quantumness is probed), we have two different expressions for discord,
\begin{eqnarray}\label{eq:D2}
D^{\leftarrow}_{A|B}(\sig_{AB}) &=& \inf_{\boldsymbol\Gamma_B^\Pi} \frac12 \ln \left(\frac{\det \sig_B \det \tilde{\boldsymbol\sigma}^{\Pi}_A}{\det \sig_{AB}}\right)\,
\nonumber \\ & & \\
D^{\rightarrow}_{A|B}(\sig_{AB}) &=&\inf_{\boldsymbol\Gamma_A^\Pi} \frac12 \ln \left(\frac{\det \sig_A \det \tilde{\boldsymbol\sigma}^{\Pi}_B}{\det \sig_{AB}}\right)\,
\nonumber
\end{eqnarray}

Discord is in fact a nonsymmetric quantity: $D^{\leftarrow}_{A|B}$ measures quantumness as revealed through the minimal disturbance induced by probing subsystem $B$, and $D^{\rightarrow}_{A|B}$ measures quantumness as revealed through the minimal disturbance induced by probing subsystem $A$. The CM ${\boldsymbol\Gamma_B^\Pi}$ (${\boldsymbol\Gamma_A^\Pi}$) denotes the seed element of the positive-operator-valued-measure Gaussian measurement $\Pi$ on subsystem $B$ ($A$), according to the characterization of \cite{giedkefiurasekdistill}, and $\tilde{\boldsymbol\sigma}^{\Pi}_A$ ($\tilde{\boldsymbol\sigma}^{\Pi}_B$) is the resulting conditional state of subsystem $A$~($B$) after the measurement $\Pi$ has been performed on $B$ ($A$). More details are available in~\cite{giordaparis,adessodatta,renyi}. Closed analytical expressions for Equations~(\ref{eq:D2}) for general two-mode Gaussian states have been derived~\cite{adessodatta,renyi}. For pure bipartite Gaussian states, $D_{A|B}^{\leftarrow}(\sig_{AB}^{\rm pure})=D_{A|B}^{\rightarrow}(\sig_{AB}^{\rm pure})=E_{A|B}(\sig_{AB}^{\rm pure})$ as~expected.

For a tripartite Gaussian state ${\op{\varrho}}_{ABC}$ with CM $\sig_{ABC}$, a measure of genuine tripartite entanglement can be defined as well from the so-called monogamy inequality \cite{ckw,contangle,3modipra,renyi}. We have
\begin{eqnarray}\label{eq:E3}
E_{A|B|C} (\sig_{ABC}) = \min \big\{\!\!\!\!&&\!\!\!\!\! E_{A|(BC)}(\sig_{ABC})-E_{A|B}(\sig_{AB})-E_{A|C}(\sig_{AC})\,\\
&&\!\!\!\!\! E_{B|(AC)}(\sig_{ABC})-E_{A|B}(\sig_{AB})-E_{B|C}(\sig_{BC})\,\nonumber\\
&&\!\!\!\!\! E_{C|(AB)}(\sig_{ABC})-E_{A|C}(\sig_{AC})-E_{B|C}(\sig_{BC})\,\big\}\nonumber\,
\end{eqnarray}
when, in particular, $\sig_{ABC}$ denotes a pure three-mode state, Equation~(\ref{eq:E3}) can be computed in closed form \cite{renyi}. We will make use of this finding to derive our results in Section~\ref{secR}. Interestingly, in such a case the residual tripartite entanglement equals the residual tripartite discord defined equivalently by replacing the $E$'s with $D^{\leftarrow}$'s in Equation~(\ref{eq:E3}) \cite{renyi}.

An even stronger indicator of nonclassical distributed correlations in a tripartite system is associated with the violation of the Svetlichny inequality \cite{svelticchio}, which reveals genuine tripartite nonlocality. Very recently, a formulation of the Svetlichny inequality has been obtained for continuous variable \mbox{states~\cite{Mauro_Referee,maurosvelt}}. Starting from the seminal observation that the Wigner function is proportional to the expectation value of displaced parity measurements \cite{wodk}, one can formulate a phase-space Bell-type inequality that is obeyed by all local hidden variable theories. The expectation value of the Svetlichny operator for a three-mode Gaussian state with CM $\sig_{ABC} \equiv \sig$ and Wigner function $W_\sig(\gr{\xi})$ defined as in Equation~(\ref{eq:wigner}), with $\gr{\xi}^{\sf T} \equiv (\gr{a},\gr{b},\gr{c})^{\sf T} \equiv (q_a,p_a,q_b,p_b,q_c,p_c)^{\sf T}$, can be written as \cite{maurosvelt}
\begin{eqnarray}\label{eq:S3}
S&=&M+M'\,, \quad \mbox{where} \\
M&=&\red{\frac{\pi^3}{8}}\big[W_\sig(\gr{a'},\gr{b},\gr{c}) +W_\sig(\gr{a},\gr{b'},\gr{c})+W_\sig(\gr{a},\gr{b},\gr{c'})-W_\sig(\gr{a'},\gr{b'},\gr{c'})
\big]\,\nonumber \\
M'&=&\red{\frac{\pi^3}{8}}\big[W_\sig(\gr{a},\gr{b'},\gr{c'}) +W_\sig(\gr{a'},\gr{b},\gr{c'})+W_\sig(\gr{a'},\gr{b'},\gr{c})-W_\sig(\gr{a},\gr{b},\gr{c})
\big]\,\nonumber
\end{eqnarray}
Here $M, M'$ are Mermin--Klyshko parameters \cite{mermin,clisho}. Violation of the Svetlichny inequality
\begin{equation}\label{eq:sven}
|S|\leq 4
\end{equation}
reveals genuine tripartite nonlocality in the state with CM $\sig$. Quantum-mechanical states can in general achieve a maximum violation of $4\sqrt2 \approx 5.65$, as it is the case for the three-qubit GHZ state \cite{svelticchio}. In the Gaussian regime, the violation of inequality (\ref{eq:sven}) has only been investigated for fully symmetric three-mode squeezed vacuum states \cite{maurosvelt} {and for partially symmetric thermal states of a three-ion setup (in the latter case, no violation was observed) \cite{Mauro_Referee}}.

\section{The System: A Driven Bose-Einstein Condensate in an Optical Ring Cavity}\label{secB}

A plethora of new research directions have arisen in cavity quantum electrodynamics once cold and ultracold atomic ensembles have been controllably placed within high-finesse optical resonators. In these systems, a global coupling scenario is realized, so that the ensemble of atoms as a whole acts onto the state of the cavity radiation field, which then reacts back on the individual atoms. There are a huge variety of settings in which entanglement between collective atomic modes and radiation fields can be produced with these setups, as very recently reviewed in \cite{2013}. Collective atomic recoil lasing, \red{predicted in \cite{bonifacio} and observed experimentally in \cite{EXPCARL}}, is the most prominent many-body instability effect occurring in a ring cavity. We study an elongated Bose-Einstein condensate, coupled to a quantized ring cavity mode, and driven by a pump laser with incident wave vector $\gr{k}$ and frequency $\omega$, far detuned from the atomic resonance frequency $\omega_0$ \cite{pariscola,pariscoladissipative}. In an approximate one-dimensional geometry, the scattered light has wave vector $\gr{k}_s \approx -\gr{k}$. In this system, the scattered cavity radiation mode and two atomic momentum side modes become macroscopically populated via a collective instability. This leads to an exponential gain in both the back-propagating radiation intensity and the atomic bunching \cite{bonifacio,pariscola,pariscoladissipative,2013}. The dimensionless interaction time can be defined as $\tau = \rho \omega_r t$, where $\omega_r=2\hbar |\gr{k}|^2/m$ is the recoil frequency, $m$ is the atomic mass, and $\rho$ is the collective atomic recoil parameter
\begin{equation}\label{eq:rho}
\rho = \left(\frac{\Omega_0}{2\Delta_0}\right)^{\frac23} \left(\frac{\omega \mu^2 N}{V \hbar \epsilon_0 \omega_r^2} \right)^{\frac13}\,
\end{equation}
where $\Omega_0 = \mu E_0/\hbar$ is the Rabi frequency of the driving laser pump (with $\mu$ the dipole matrix element and $E_0$ the electric field amplitude), $\Delta_0 = \omega - \omega_0$ is the pump-atom detuning, $N$ is the number of atoms in the cavity mode volume $V$, and $\epsilon_0$ is the vacuum permittivity (see \cite{bonifacio,pariscola,pariscoladissipative} for details).

We assume that the atoms are delocalized inside the condensate and that, at zero temperature, the momentum uncertainty can be neglected compared with $2\hbar |\gr{k}|$. This approximation is valid when the longitudinal size of the condensate is much larger than the wavelength of the incident radiation. The momentum levels of the atomic ensemble are quantized; here we will treat all the strongly populated momentum modes classically and all weakly populated modes quantum mechanically, as in the standard undepleted pump approximation for nonlinear optics \cite{2013}. In these conditions, the equilibrium state with no cavity field consists of all the atoms in the same initial state $\ket{n_0}$, \emph{i.e}., all the atoms moving with the same momentum $2n_0\hbar |\gr{k}|$. In the linear regime around the equilibrium atomic level $\ket{n_0}$, only two momentum side levels will be populated. The system can be thus described in terms of three coupled harmonic oscillators (bosonic modes). Modes $1$ and $2$ are collective atomic modes corresponding to atoms being respectively in the momentum level $\ket{n_0-1}$ and $\ket{n_0+1}$, \emph{i.e}., having respectively lost and gained a quantum recoil momentum $2 \hbar |\gr{k}|$ due to the two-photon Bragg scattering between the driving laser and the cavity field. Mode $3$ describes the scattered cavity field light, with momentum $\gr{k_s}$ and frequency $\omega_s = c |\gr{k_s}|$. In the limit of a high-finesse, lossless cavity, the evolution is nondissipative and three-mode system remains in a pure Gaussian state under the unitary dynamics induced by the~Hamiltonian
\begin{equation}\label{eq:HAM}
\op{H}_{123} = \delta_+ \op{b}^{\dag}_2 \op{b}_2 - \delta_- \op{b}^{\dag}_1 \op{b}_1 +i \sqrt{\frac{\rho}{2}}\left[\big(\op{b}^{\dag}_1+\op{b}_2\big)\op{b}_3^{\dag} - \big(\op{b}_1+\op{b}_2^{\dag}\big)\op{b}_3\right]\,
\end{equation}
where $\delta_\pm = \delta \pm 1/\rho$, $\delta=\Delta+2n_0/\rho$, and $\Delta=(\omega-\omega_s)/(\rho \omega_r)$ is the detuning between the driving pump laser and the cavity field. The interested reader can find relevant details on the derivation of Equation~(\ref{eq:HAM}) and on the general description of the system in~\cite{bonifacio,pariscola,pariscoladissipative,2013}. Fixing the detuning so that $\delta=1/\rho$ \cite{pariscoladissipative}, the solution to the dynamical equations, which display collective atomic recoil lasing instability, will depend on just two parameters, the recoil factor $\rho$, Equation~(\ref{eq:rho}), and the dimensionless time $\tau$. Two relevant regimes of operation can be realized, namely the quantum ($\rho < 1$) and semiclassical ($\rho \gg 1$) good-cavity regimes \cite{pariscoladissipative}. For any $\rho$ and $\tau$, the elements of the $6\times 6$ covariance matrix $\sig_{123}(\rho,\tau)$ of the three-mode system can be calculated analytically from the implicit expression
\[\sum_{j=1}^3 |\Lambda_j|^2 = \frac12\left(\gr{x}^{\sf T}\ \sig_{123}\ \gr{x}\right)\,\] where $\gr{x}^{\sf T}=(q_1,p_1,q_2,p_2,q_3,p_3)$, $\Lambda_1 = g_1 z_1 - h_1 z_1^*$, $\Lambda_2=-g_2^*z_1^*+h_2^*z_2+f_2^*z_3$, $\Lambda_3=-g_3^*z_1^*+h_3^*z_2+f_3^*z_3$, $z_j=\frac{1}{\sqrt{2}}(p_j-i q_j)$, and the functions $f_j, g_j, h_j$, which depend on $\rho$ and $\tau$, are given explicitly in the Appendix of \cite{pariscola}. Initially, at $\tau=0$, the system is in the vacuum, \emph{i.e}., in a factorized pure state of the three modes, so that $\sig_{123}(\rho, 0) = \id$. As soon as $\tau>0$, the three modes become populated and multipartite quantum correlations arise among them, as described in the next section.

\section{The Results: Tripartite Entanglement, Quantumness and Nonlocality}\label{secR}

In~\cite{pariscola,pariscoladissipative}, the authors analyzed the entanglement properties of the three-mode system in the state characterized by the CM $\sig_{123}$. By classifying its separability, they found that such a state is fully inseparable in the whole relevant parameter range ($\rho,\tau>0$). That is, every mode $j$ is entangled with the group formed by the other two modes. The reduced two-mode partitions of $\sig_{123}$, however, are not all entangled. In particular, no entanglement is ever created between modes $2$ and $3$, as it can be checked by the partial transposition criterion \cite{ourreview,pariscola}. Recall that modes $1$ and $2$ are the lower and upper atomic side momentum levels, while mode $3$ is the cavity field mode. \red{Physically, this can be understood by looking at the Hamiltonian in Equation~(\ref{eq:HAM}). The interaction between the light mode $3$ and the atomic mode $1$ is of the parametric down-conversion type, \emph{i.e}.,~an active operation that induces two-mode squeezing and entanglement between these two modes. Conversely, the interaction between modes $2$ and $3$ amounts to a passive beamsplitter-type operation, which does not create entanglement when the two inputs are thermal states, as in the present case.}

Here, we are able to calculate analytically the degree of bipartite entanglement in the various reduced bipartitions, as well as the degree of genuine tripartite entanglement established between the atomic modes $1$ and $2$ and the light mode $3$ by collective atomic recoil. We will omit the explicit expressions as they are too cumbersome to reveal any insight, and we will resort to a graphical presentation of our findings. It suffices to recall that, in a pure three-mode Gaussian state such as that described by $\sig_{123}$, all the relevant informational and correlation properties depend only on the three local symplectic invariants $\det(\sig_j)$ \cite{3modipra}. In the considered system, the three local determinants are in general all different, which means that we have a completely nonsymmetric state; however, in our case these quantities satisfy the constraint $\sqrt{\det{\sig_1}}-\sqrt{\det{\sig_2}}-\sqrt{\det{\sig_3}}+1=0$. The latter arises since the operator $\op{n}_1-\op{n}_2-\op{n}_3$ is a constant of motion for our system described by the Hamiltonian in Equation~(\ref{eq:HAM}), and $\det(\sig_j)=(2 \langle \op{n}_j \rangle+1)^2$. Simple limiting expressions for the mode populations $\langle \op{n}_j \rangle$, valid approximately in the quantum ($\rho < 1$) or semiclassical ($\rho \gg 1$) good-cavity regimes at $\tau \gg 0$, are reported in \cite{pariscoladissipative}.

\begin{figure}[t!]
\centering
\subfigure{
\includegraphics[height=5.5cm]{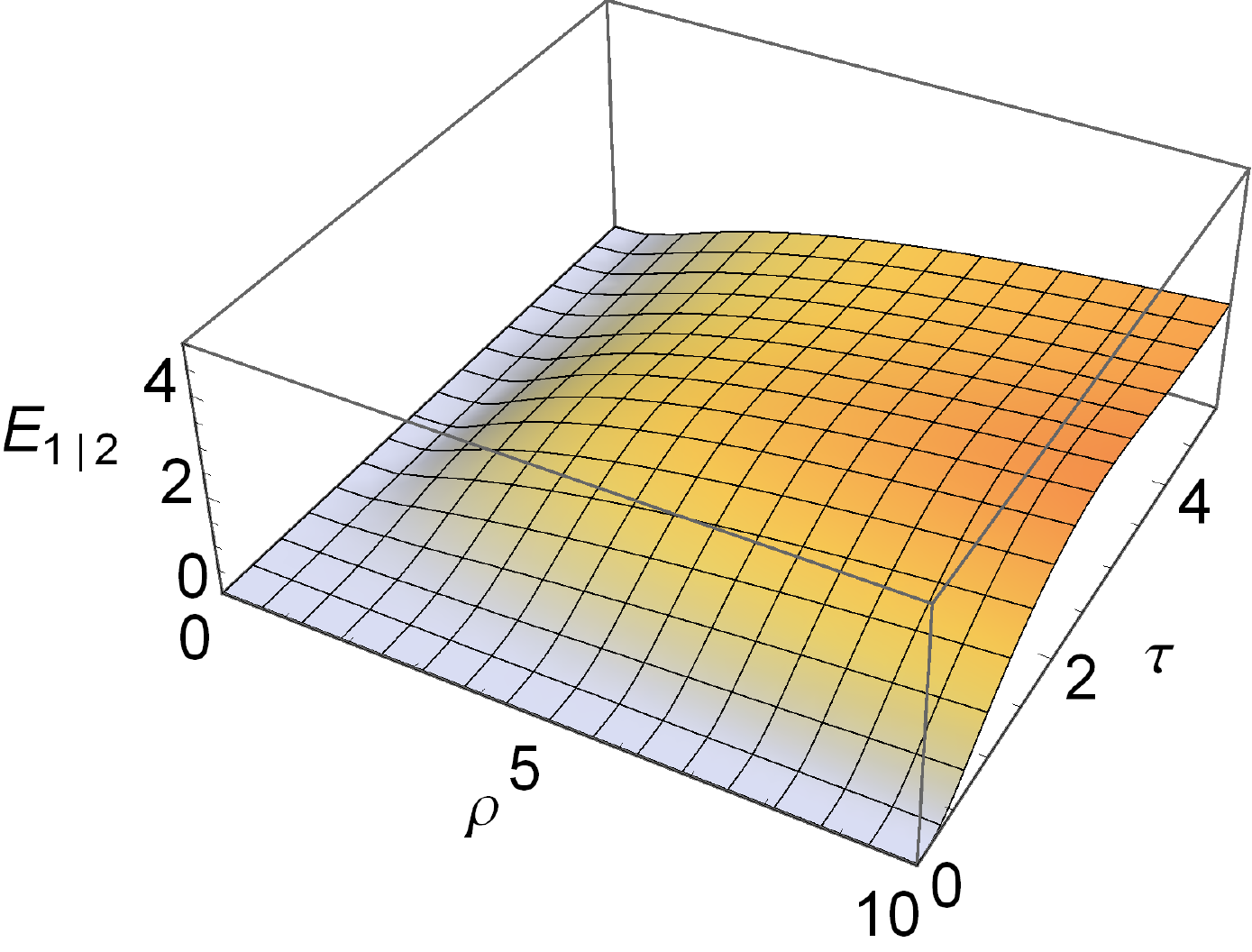}}\hspace*{1cm}
\subfigure{
\includegraphics[height=5.5cm]{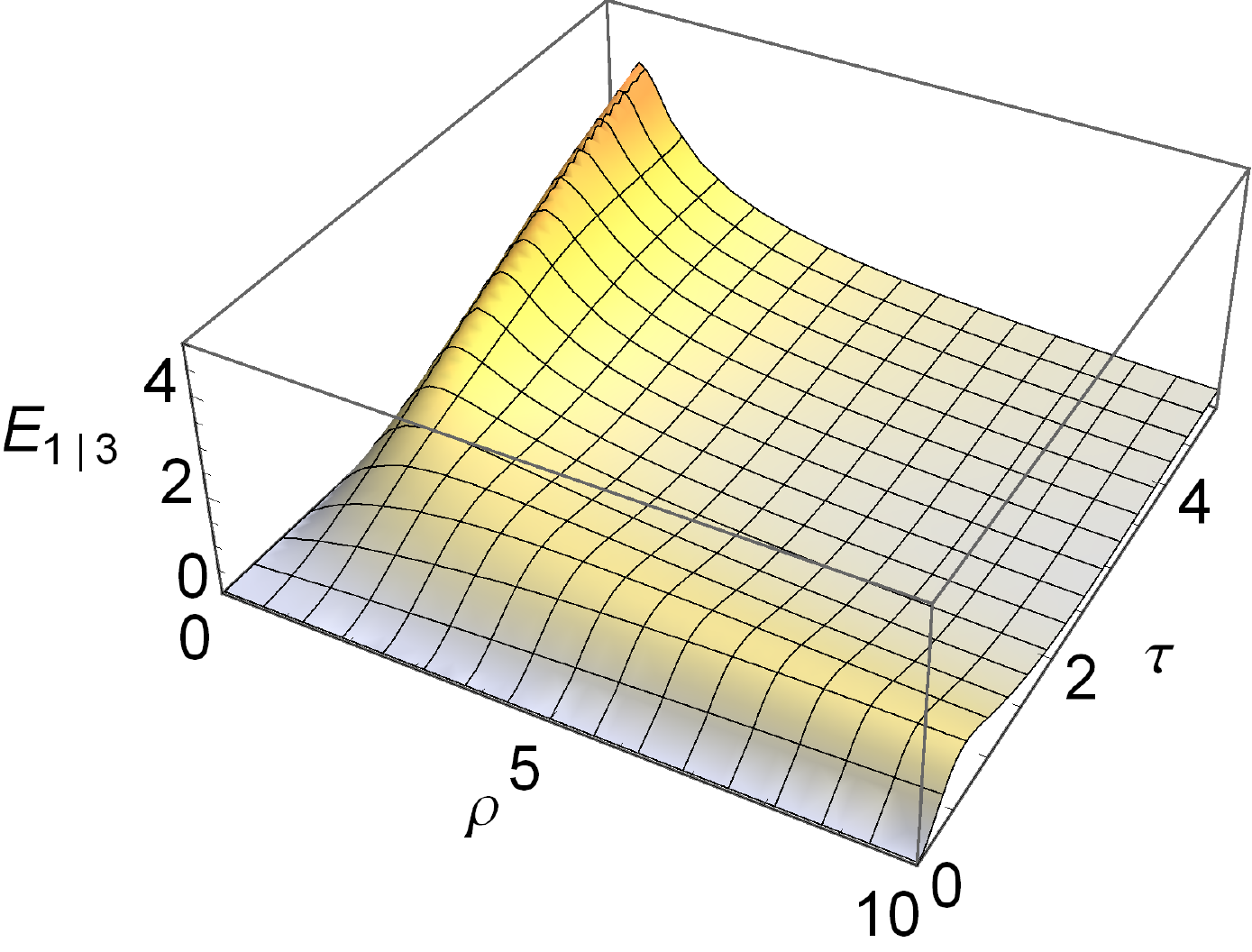}}
\caption{Two-mode entanglement in the reduced bipartite states $\sig_{12}$ (atom-atom, left) and $\sig_{13}$ (atom-light, right) of the three-mode system described by the Hamiltonian (\ref{eq:HAM}), plotted \emph{versus} the recoil parameter $\rho$ and the dimensionless time $\tau$.}
\label{figE2}
\end{figure}

In Figure~\ref{figE2} we plot the reduced two-mode entanglement between modes $1$ and $2$, $E_{1|2}(\sig_{12})$ (left), and between modes $1$ and $3$, $E_{1|3}(\sig_{13})$ (right), as calculated from Equation~(\ref{eq:GR2_ent}) \cite{renyi}. We notice how in the quantum good-cavity regime $(\rho<1)$ a strong atom-field entanglement $E_{1|3}$ is engineered by the dynamics. On the contrary, in the semiclassical good-cavity regime $(\rho \gg 1)$, a direct atom-atom entanglement $E_{1|2}$, mediated by the field, prevails instead. Proposals to exploit the atom-light entanglement $E_{1|3}$ in the quantum regime for a hybrid teleportation were discussed in \cite{pariscola}.

\begin{figure}[t!]
\centering
\subfigure{
\includegraphics[height=5.5cm]{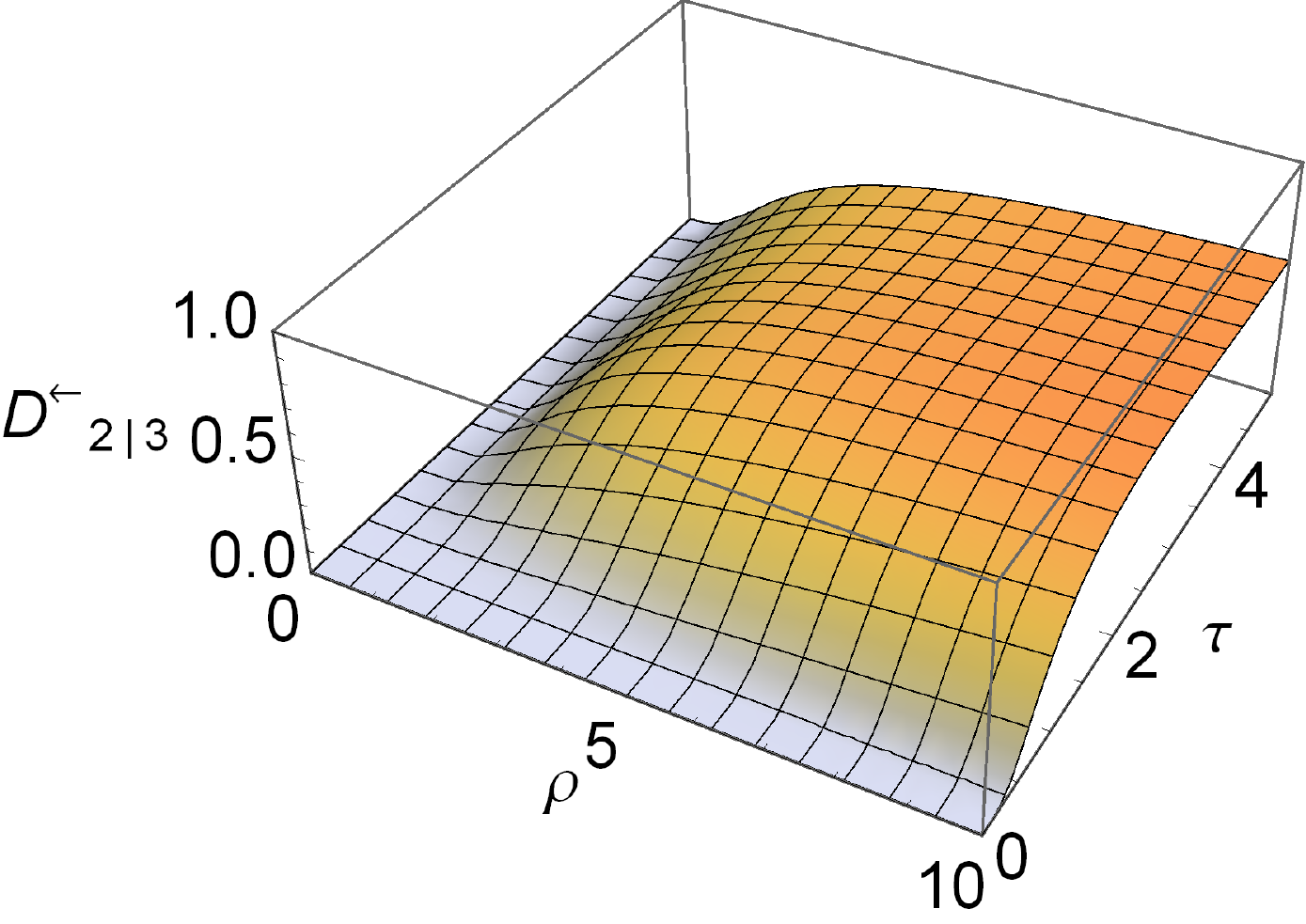}}\hspace*{1cm}
\subfigure{
\includegraphics[height=5.5cm]{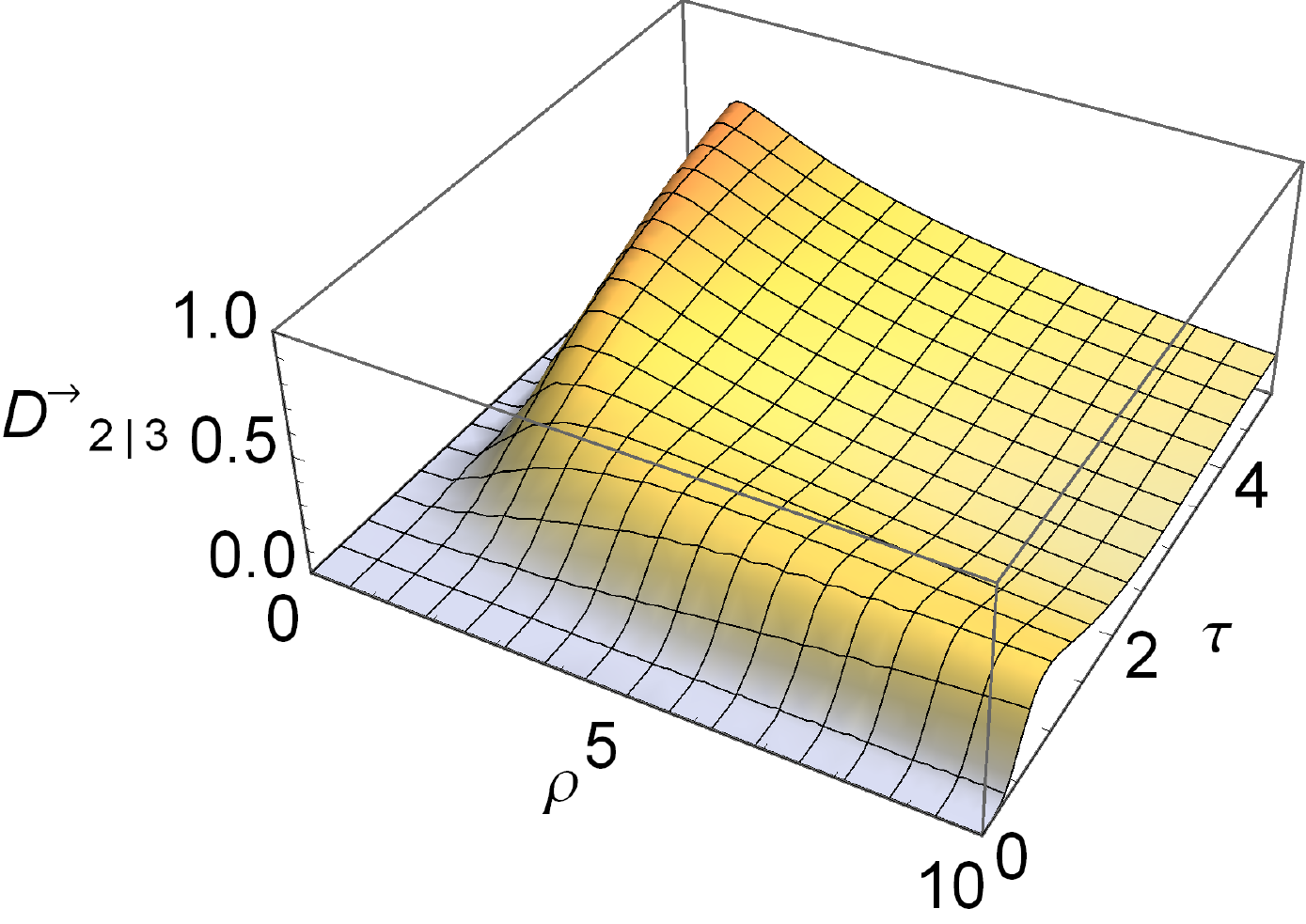}}
\caption{Two-mode discord in the reduced atom-light bipartite state $\sig_{23}$ of the three-mode system described by the Hamiltonian (\ref{eq:HAM}), revealed by probing mode $3$ (left) or mode $2$ (right), plotted \emph{versus} the recoil parameter $\rho$ and the dimensionless time $\tau$.}
\label{figD2}
\end{figure}

We mentioned previously that the other atom-field partition $\sig_{23}$ is always separable. It is natural to question whether such a reduced system still displays nonclassical correlations more general than entanglement, in the form of quantum discord \cite{zurek,modireview}. In Figure~\ref{figD2} we plot the discord between modes~$2$ and $3$, calculated from Equation~(\ref{eq:D2}), as revealed by probing the light mode $3$, $D^{\leftarrow}_{2|3}(\sig_{23})$ (left), and by probing the atomic mode $2$, $D^{\rightarrow}_{2|3}(\sig_{23})$ (right). We observe, interestingly, that $D^{\rightarrow}_{2|3}$ behaves very similarly to the atom-field entanglement $E_{1|3}$, while $D^{\leftarrow}_{2|3}$ behaves very similarly to the atom-atom entanglement $E_{1|2}$. Both discords, in their appropriate regimes, converge to $\ln2$ for $\tau \gg 0$, which is the maximum allowed value for a Gaussian separable state \cite{adessodatta,renyi}. This study reveals that the quantumness of correlations in the partition $2|3$ is always significant and best revealed by measuring the atomic mode~$2$ if we are in the quantum regime of small $\rho$, or the light mode $3$ if we are instead in the semiclassical regime of large $\rho$.

\begin{figure}[t!]
\centering
\subfigure{
\includegraphics[height=5.5cm]{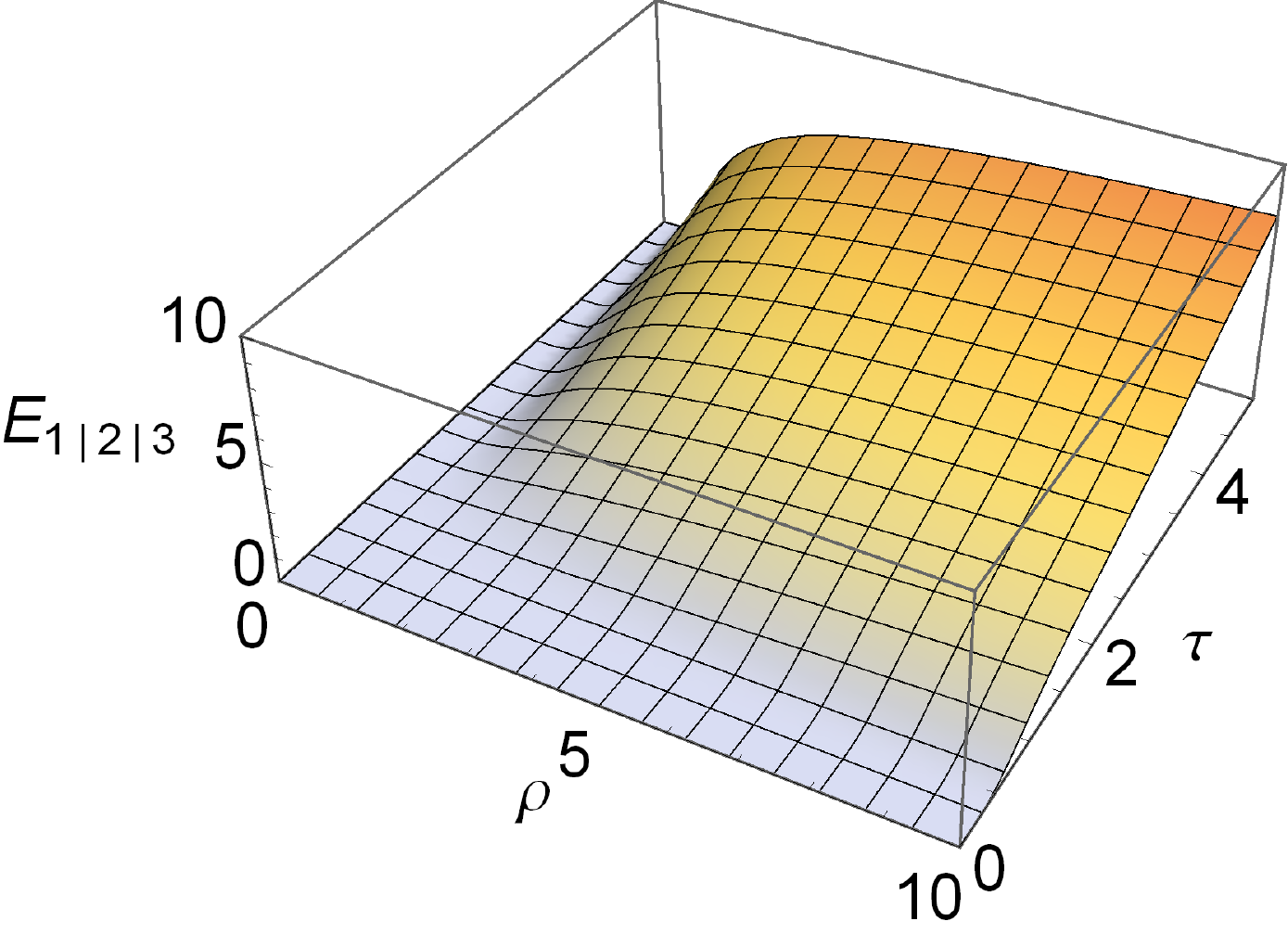}}\hspace*{1cm}
\subfigure{
\includegraphics[height=5.5cm]{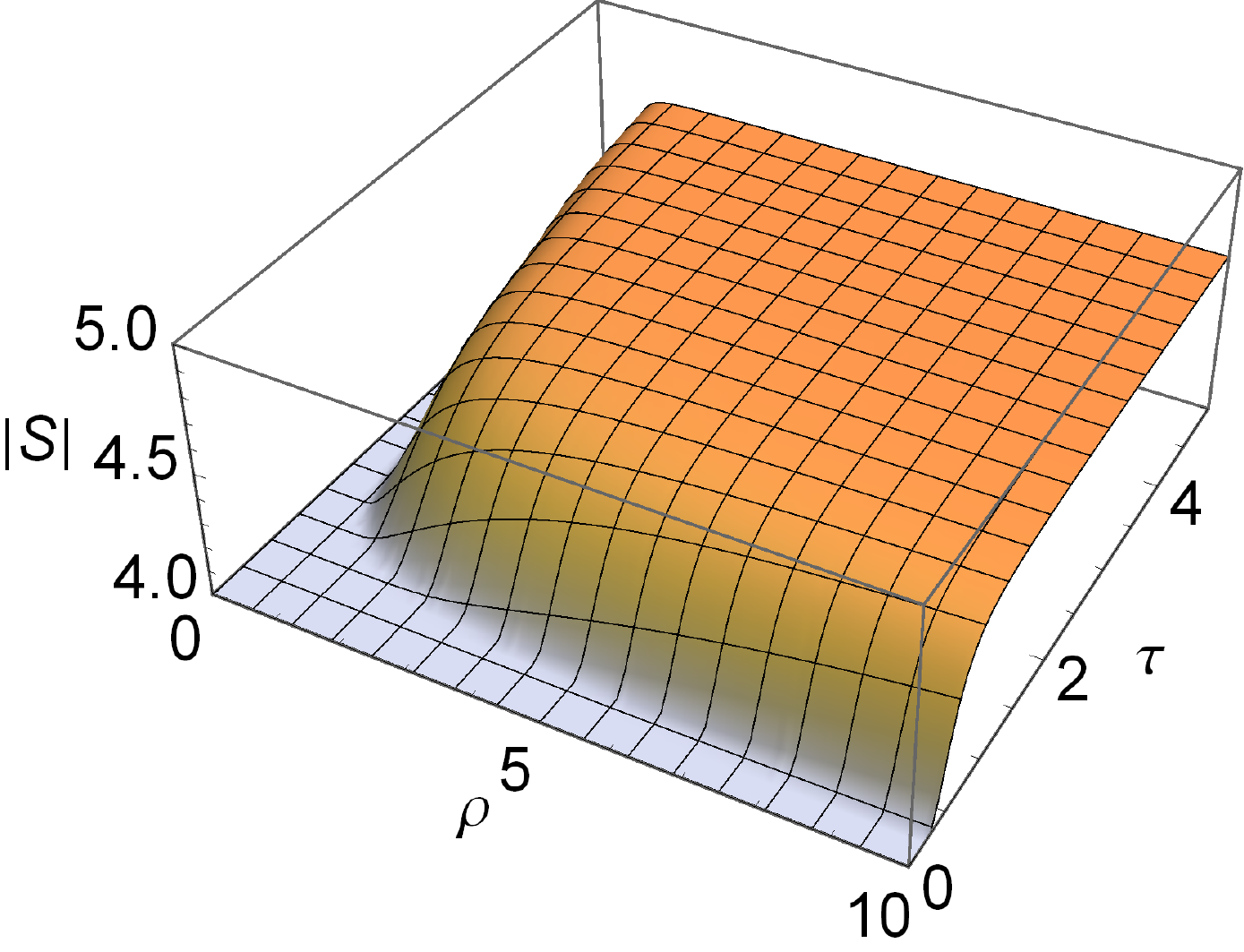}}
\caption{Genuine tripartite entanglement (left) and maximal violation of the Svetlichny inequality revealing genuine tripartite nonlocality (right) for the state $\sig_{123}$ of the three-mode system described by the Hamiltonian (\ref{eq:HAM}), plotted \emph{versus} the recoil parameter $\rho$ and the dimensionless time $\tau$.}
\label{fig3}
\end{figure}

We are now in a position to analyze and quantify the genuine multipartite entanglement among the three modes $1$, $2$ and $3$ established via the dynamics generated by the Hamiltonian of Equation~(\ref{eq:HAM}). We find that the minimum in Equation~(\ref{eq:E3}) is attained by the decomposition that involves mode $1$ as the probe mode, so that
\begin{equation}\label{eq:E3R}
E_{1|2|3} (\sig_{123}) = E_{1|(23)}(\sig_{123})-E_{1|2)}(\sig_{12})-E_{1|3}(\sig_{13})\,
\end{equation}
The tripartite entanglement $E_{1|2|3}$ is plotted in Figure~\ref{fig3} (left) as a function of $\rho$ and $\tau$. We observe a strong tripartite entanglement that grows unboundedly with both increasing parameters and is significantly larger than the bipartite entanglements in any reduced bipartition. The instability induced by the collective atomic recoil lasing mechanism can be thus viewed as a useful source of genuine three-mode entanglement in an ultracold-atom-light interfaced setup, which can be exploited for instance to implement hybrid teleportation networks \cite{telenet}. It is in order to remark that numerical analysis in the non-ideal case of a dissipative, lossy cavity, and in presence of atomic decoherence, reveals that the full inseparability of the systems is preserved \cite{pariscola,pariscoladissipative}, meaning that the tripartite entanglement is robust against realistic imperfections. \red{Specifically, taking into account cavity losses and atomic decoherence as modelled in \cite{pariscoladissipative}, the state of the system in our regime of consideration can still be described as a Gaussian, yet mixed, three-mode state. All the tools adopted in this paper can still be used in such a more general case, although the genuine tripartite entanglement $E_{1|2|3}$ requires typically a numerical optimization to be computed. This lies outside the scope of the present paper. To give a flavour of what is achievable in laboratory, let us recall that in the experimental demonstration of \cite{EXPCARL}, a high-$Q$ cavity was used with amplitude decay time of about $3$ $\mu$s, while collective atomic recoil lasing effects were detected up to more than $1.5$ ms; atomic collisions were negligible in that implementation.}

We finally enquire whether the genuine tripartite nonlocality of the considered three-mode system, a stronger feature than genuine tripartite entanglement, could also be revealed. This would manifest in a violation of the Svetlichny inequality (\ref{eq:sven}). We have numerically optimized the Svetlichny parameter~$|S|$, Equation~(\ref{eq:S3}), over the settings $\gr{\xi}, \gr{\xi'}$ for the atom-atom-field three-mode state $\sig_{123}$ of our system. The resulting maximal value of~$|S|$ is plotted in Figure~\ref{fig3} (right) as a function of $\rho$ and $\tau$. Interestingly, although the state is fully inseparable as soon as $\tau>0$, the inequality (\ref{eq:sven}) is not violated for a small but finite range of values of $\tau, \rho \ll 1$, in which $|S|=4$. For sufficiently large values of $\rho$ and $\tau$, a violation is observed, which grows monotonically with the genuine tripartite entanglement $E_{1|2|3}$. Remarkably, the optimal violation rapidly saturates to the limiting value $\frac{16}{3^{\frac98}} \approx 4.65$, which appears to be the maximum allowed value for any three-mode Gaussian state \cite{inprep}. Therefore the considered driven intracavity condensate system constitutes a valuable setup to verify experimentally a strong violation of nonlocal realism by phase-space measurements in a genuine multipartite setting, on top of the presence of genuine multipartite entanglement and quantum correlations in general.

\section{Conclusions}\label{secC}

In this special issue contribution we have investigated entanglement, general quantum correlations and genuine multipartite nonlocality in the dynamics of a three-mode Gaussian state, corresponding to a Bose-Einstein condensate interacting with a cavity field in the presence of a strong off-resonant pumping laser \cite{pariscola}. In the good-cavity limit, the mechanism of collective atomic recoil \cite{bonifacio,EXPCARL} gives rise to the build of genuine tripartite entanglement shared among two atomic momentum side modes and the cavity field mode, which increases unboundedly with the evolution time. The reduced two-mode bipartitions display quantum correlations, in the form of entanglement \cite{entanglement} for two of them, and discord without entanglement \cite{modireview} for the third one. The considered setup allows to reveal genuine tripartite nonlocality via a violation of the Svetlichny inequality on local realism \cite{svelticchio,maurosvelt}, in a broad range of the physical~parameters.

The value of this paper lies not only in the comprehensive analysis of the specific Bose-Einstein condensate system we investigated, which is of relevance for the implementation and control of quantum information processing in atom-light interfaced setups \cite{qinternet,2013}, but also in the concise introduction to the up-to-date mathematical formulation of nonclassical features of Gaussian states in continuous variable systems \cite{ourreview,pirandolareview,renyi}. The adopted measures based on the R\'enyi entropy of order $2$ provide a sound and flexible toolbox for the study of a variety of physical systems. With little alterations, the analysis presented in this paper could be adapted to study, e.g., ultracold atoms coupled to optomechanical devices, ion crystals, mechanical membranes inside a cavity, as well as of course all-optical setups \cite{book}. This paper also presents the first instance of a nonsymmetric Gaussian state exhibiting a violation of the Svetlichny inequality, generalizing the scope of the examples reported in \cite{Mauro_Referee,maurosvelt}.

A detailed investigation of genuine tripartite nonlocality for arbitrary pure and mixed three-mode Gaussian states will be the subject of a further study \cite{inprep}.

\newpage

\acknowledgements{Acknowledgements}
We acknowledge the University of Nottingham for financial support via a Nottingham Advance Research Fellowship (S.P.) and a Research Development Fund grant (G.A.). We thank Matteo G. A. Paris for fruitful discussions.



\end{document}